\documentclass[sigconf,nonacm]{acmart}
\acmConference[ESEC/FSE 2023]{The 31st ACM Joint European Software Engineering Conference and Symposium on the Foundations of Software Engineering}{11 - 17 November, 2023}{San Francisco, USA}

\usepackage[switch,columnwise]{lineno}

\usepackage{caption,subcaption}
\usepackage{soul}
\usepackage{xspace}    
\usepackage{booktabs}
\usepackage{cleveref}
\usepackage{graphicx}
\graphicspath{{./figures/}}
\usepackage{textcomp}
\usepackage{xcolor}
\usepackage{array}

\newcommand{\FuncTime}{30 seconds\xspace}
\newcommand{\FuncTol}{$3\frac{km}{h}$}
\newcommand{\LongAccFor}{$5\frac{m}{s^{2}}$}
\newcommand{\LongAccBack}{$-3.5\frac{m}{s^{2}}$}
\newcommand{\PitchAcc}{$3\frac{rad}{s^{2}}$}
\newcommand{\JerkThres}{$10\frac{m}{s^{3}}$}

\AtBeginDocument{\providecommand\BibTeX{{\normalfont B\kern-0.5em{\scshape i\kern-0.25em b}\kern-0.8em\TeX}}}

\newcommand{\NAME}{\textsc{HECATE}\xspace}

\usepackage{graphbox}
\usepackage{ifthen}
\usepackage{mathtools}
\usepackage[most]{tcolorbox}
\usepackage{fancybox}
\usepackage{makecell}
\usepackage{multirow}
\usepackage{booktabs}

\usepackage{caption}
\usepackage{subcaption}
\usepackage{epstopdf}

\usepackage{tcolorbox}
\usepackage{xspace}
\usepackage[switch]{lineno}

\usepackage[binary-units,per-mode=symbol,detect-weight=true, detect-family=true,inter-unit-product =\cdot]{siunitx}

\usepackage{algpseudocode}
\usepackage{algorithm}

\newcommand{\developmentTime}{300 hours}

\usepackage[framemethod=TikZ]{mdframed}

\newenvironment{Answer}[1][]{\ifstrempty{#1}{\mdfsetup{frametitle={\tikz[baseline=(current bounding box.east),outer sep=0pt]
      \node[line width=0pt,anchor=east,rectangle,draw=white,fill=white]
    ;}}
  }{\mdfsetup{frametitle={\tikz[baseline=(current bounding box.east),outer sep=0pt]
      \node[anchor=east,rectangle,draw=white,fill=white]
    {\strut #1};}}}\mdfsetup{innertopmargin=-5pt,linecolor=black,linewidth=0.5pt,topline=true,frametitleaboveskip=\dimexpr-\ht\strutbox\relax,}
  \begin{mdframed}[]\relax }{\end{mdframed}}

\newcommand{\phase}[1]{\tikz[baseline=(X.base)]\node [draw=black,fill=white,thick,rectangle,inner sep=1.5pt, rounded corners=2pt](X){\color{black}\textbf{#1}};}

\newcommand{\simulink}{Simulink\textsuperscript{\tiny\textregistered}\xspace}

\newcommand{\system}{\ensuremath{S}\xspace}
\newcommand{\fitness}{\ensuremath{f}\xspace}
\newcommand{\asmpt}{\ensuremath{A}\xspace}
\newcommand{\budget}{\ensuremath{T}\xspace}
\newcommand{\requirement}{\ensuremath{\varphi}}
\newcommand{\inputs}{\ensuremath{\texttt{i}}\xspace}

\newcommand{\totalModelVersions}{21\xspace}
\newcommand{\totalFailure}{14\xspace}
\newcommand{\funcFailure}{6\xspace}
\newcommand{\drivFailure}{3\xspace}
\newcommand{\funcdrivFailure}{5\xspace}

\newcommand{\percentageFailure}{66.7$\%$\xspace}
\newcommand{\percentageFailureFunc}{42.9$\%$\xspace}
\newcommand{\percentageFailureDriv}{21.4$\%$\xspace}
\newcommand{\percentageFailureFuncDriv}{35.7$\%$\xspace}

\newcommand{\NBlocksCC}{322\xspace}

\newcommand{\itmin}{1\xspace}
\newcommand{\itmax}{14\xspace}
\newcommand{\itstd}{3.9\xspace}
\newcommand{\itavg}{3.8\xspace}

\newcommand{\timemin}{48.2s\xspace}
\newcommand{\timemax}{784.5s\xspace}
\newcommand{\timestd}{262.2s\xspace}
\newcommand{\timeavg}{245.9s\xspace}

\newcommand{\simulationtime}{100s\xspace}
\newcommand{\maxiterationsUR}{20\xspace}

\newcommand{\maxsimulationtimeUR}{34min\xspace}
\newcommand{\maxiterationsSA}{50\xspace}

\newcommand{\VICRTBlocks}{1125\xspace}

\newcommand{\totalBlocks}{1461\xspace}

\usepackage{pifont}
\newcommand{\cmark}{\ding{51}}
\newcommand{\xmark}{\ding{55}}

\newcommand{\STaLiRo}{S-TaLiRo\xspace}
\newcommand{\PsiTaliro}{\ensuremath{\Psi}-TaLiRo\xspace}
\newcommand{\Breach}{Breach\xspace}
\newcommand{\ATheNA}{ATheNA\xspace}

\newcommand{\STGEM}{STGEM\xspace}
\newcommand{\ARIsTEO}{ARIsTEO\xspace}
\newcommand{\FalStar}{\textsc{FalStar}\xspace}
\newcommand{\foresee}{\textsc{ForeSee}\xspace}

\newcommand{\falsify}{falsify\xspace}

\newcommand{\FalCAuN}{\texttt{FalCAuN}\xspace}

\definecolor{keywordcolor}{HTML}{1e46be}
\definecolor{stepcolor}{HTML}{724722}

\newcommand{\lit}[1]{\textbf{\texttt{\textcolor{keywordcolor}{#1}}}}

\newcommand{\simulinkvariable}[1]{\text{#1}}

\newcommand{\TrueValue}{\textsc{true}\xspace}
\newcommand{\FalseValue}{\textsc{false}\xspace}

\newcommand{\step}[1]{\textbf{\textbf{\color{stepcolor}#1}}\xspace}

\usepackage{amsmath}

\newcommand{\SLTest}{Simulink\textsuperscript{\tiny\textregistered}
Test\textsuperscript{\tiny TM}\xspace}

\begin{document}

\title{Test Case Generation for Drivability Requirements of an Automotive Cruise Controller:\\
An Experience with an Industrial Simulator}

\author{Federico Formica}
\affiliation{\institution{McMaster University}
    \city{Hamilton}
    \country{Canada}
}

\author{Nicholas Petrunti}
\affiliation{\institution{McMaster University}
    \city{Hamilton}
    \country{Canada}
}

\author{Lucas Bruck}
\affiliation{\institution{McMaster University}
    \city{Hamilton}
    \country{Canada}
}

\author{Vera Pantelic}
\affiliation{\institution{McMaster University}
    \city{Hamilton}
    \country{Canada}
}

\author{Mark Lawford}
\affiliation{\institution{McMaster University}
    \city{Hamilton}
    \country{Canada}
}

\author{Claudio Menghi}
\affiliation{\institution{University of Bergamo}
    \city{Bergamo}
    \country{Italy}
}
\affiliation{\institution{McMaster University}
    \city{Hamilton}
    \country{Canada}
}

\renewcommand{\shortauthors}{Formica, et al.}

\begin{abstract}
Automotive software development requires engineers to test their systems to detect violations of both functional and drivability requirements.
Functional requirements define the functionality of the automotive software.
Drivability requirements refer to the driver's perception of the interactions with the vehicle; for example, they typically require limiting the acceleration and jerk perceived by the driver within given thresholds. 
While functional requirements are extensively considered by the research literature,  drivability requirements garner less attention.

This industrial paper describes our experience assessing the usefulness of an automated search-based software testing (SBST) framework in generating failure-revealing test cases for functional and drivability requirements. 
Our experience concerns the VI-CarRealTime simulator, an industrial virtual modeling and simulation environment widely used in the automotive domain.
We designed a Cruise Control system in \simulink for a four-wheel vehicle, in an iterative fashion, by producing \totalModelVersions model versions. 
We used the SBST framework for each version of the model to search for failure-revealing test cases revealing requirement violations.
Our results show that the SBST framework successfully identified a failure-revealing test case  for \percentageFailure of our model versions, 
requiring, on average,  \timeavg and \itavg iterations.
We present lessons learned, reflect on the generality of our results, and discuss how our results improve the state of practice. 
\end{abstract}


\maketitle

\section{Introduction}
\label{sec:Intro}
\emph{Search-based software testing} (SBST) is a technique used to automatically generate test cases that show the violation of the system requirements~\cite{harman2001search}.
SBST frameworks support software development in many domains, such as real-time, concurrent, distributed, embedded, and safety-critical~\cite{ali2009systematic}.
SBST is also extensively used in the context of cyber-physical systems (CPS), including automotive applications (e.g.,~\cite{stocco2023model,liu2019effective,9869302,jahangirova2021quality,arrieta2016test}). 
Despite their extensive adoption, the usefulness of SBST depends on the assumptions and peculiarities of the different domains since the types of requirements under consideration usually depend on the application domains, e.g., slow response time is a fault relevant for real-time systems~\cite{ali2009systematic}, but not necessarily for others.

This work focuses on the automotive domain and specifically on \emph{Cruise Controller} (CC) development.
CC is a software component that regulates the vehicle's speed. Their global market was valued at USD 34.7 billion in 2022 and is expected to grow to USD 65 billion by 2032~\cite{MarkedCruiseControl}, driven by the increasing demand for safety and comfort, the technology advancements, and the rise of autonomous cars~\cite{koopman2016challenges}.
CC needs to satisfy functional requirements while ensuring the comfort of the driver~\cite{althoff2020provably,bjornander2008adaptive}.
\emph{Drivability requirements} specify properties related to the comfort of the driver and pilot the CC development~\cite{boschCC,althoff2020provably,opila2011energy,Bruck2022,barroso2023driver,9490149,drivability}. 

Testing is one of the techniques used to search for violations of the CC requirements~\cite{8952365,barthauer2019testing,barnes2020scalable,lin2020safety,mehra2015adaptive}.
The assessment of the usefulness of testing techniques in practice via empirical studies is fundamental for the academic and industrial communities, which need experimental evaluation results to drive their (business) decisions and understand the advantages and limitations of SBST support for CC development.
However, for CC development, most of the studies publicly available in the research literature (e.g.,~\cite{koschi2019computationally,8952365,tuncali2019requirements,7795751,
calo2020simultaneously,laurent2022parameter})
focus on functional requirements and do not empirically assess the SBST frameworks by considering the versions of the model produced during the incremental and iterative development of  CC. 
This paper addresses this limitation by empirically assessing the usefulness of SBST in detecting failure-revealing test cases for drivability requirements during the end-to-end development of a CC for an industrial simulator.

Our \emph{case study} is the VI-CarRealTime (VICRT) Simulator~\cite{vicar}, an industrial real-time simulation environment widely used by automotive companies (e.g., Brembo~\cite{Brembo}). 
VICRT enables engineers to develop control software for their vehicles.
It integrates with well-known development tools for cyber-physical system design, such as Matlab/\simulink~\cite{Simulink},
and supports both software-in-the-loop (SIL) and hardware-in-the-loop (HIL) simulations. 
We considered \emph{\simulink} for developing the CC since it  is a widely used framework for developing control systems in the industry~\cite{boll2021characteristics,jaskolka2021repository}. 
For the assessment of SBST, we use \SLTest~\cite{SimulinkTest} for test case definition, since it is a standard tool for test case specification in \simulink.
\SLTest enables engineers to specify test cases using \simulink Test Sequence~\cite{TestSequence} and Test Assessment~\cite{TestAssessment}  blocks, also called for short Test Blocks in this work.
\emph{Test Sequences}  specify  test inputs, while 
\emph{Test Assessments} specify the procedure to check the system requirements. 
We selected \NAME~\cite{Hecate} as our SBST framework since it generates test cases in collaboration with Test Blocks.

We iteratively and incrementally developed a complex CC for our industrial simulator.
The development activity required approximately 300 hours spread over eight months, during which we produced \totalModelVersions versions of the CC referring to seven major versions.
We extensively used the SBST framework during the CC development by running SIL and HIL experiments,
and assessed the usefulness of SBST by measuring its effectiveness, i.e., how helpful SBST is in detecting model failures, and efficiency, i.e., the time required for detecting the failures.
Our results show that the SBST framework successfully identified a failure-revealing test case  for \percentageFailure of our model versions (\totalFailure out of \totalModelVersions).
The SBST framework required  \timeavg and \itavg iterations (on average) to detect the failure-revealing test cases.
Only the last version of the model can pass the most complex and generic scenario we tested.

In summary, this work addresses the following problems: 
\begin{itemize}
    \item[\textbf{P1}] it assesses the usefulness of SBST in detecting failure-revealing test cases for the development of a complex CC for an \emph{industrial simulator};
    \item[\textbf{P2}] it assesses the usefulness of SBST in detecting failure-revealing test cases for \emph{drivability requirements}, an important and large category of automotive requirements;   
    \item[\textbf{P3}] it assesses the usefulness of SBST driven by \emph{\simulink Test Blocks}, that are standard tools for test case specification in \simulink.
\end{itemize}
These research problems are relevant to the industry and motivated by forthcoming industrial challenges. Understanding the usefulness of SBST in detecting failure-revealing test cases for CC development~(\textbf{P1}) enables automotive industries to assess how beneficial SBST techniques are and helps them to evaluate how to use them within their development processes.
Understanding the usefulness of SBST in detecting failure-revealing test cases for drivability requirements for the CC~(\textbf{P2}) enables automotive industries to know how frequently these requirements are violated compared to the functional requirements.
This information helps them understand how to prioritize and analyze different requirements across the CC development.
Finally, understanding the usefulness of SBST driven by \simulink Test Blocks~(\textbf{P3}) helps automotive industries assess how beneficial this technique is in developing complex automotive systems.

This work is organized as follows.  
Section~\ref{sec:caseStudy} describes our automotive case study.
Section~\ref{sec:DevCC} summarizes the CC development activities and the application of the testing framework.
Section~\ref{sec:sbst} presents SBST and \NAME, the SBST framework considered in this work.
Section~\ref{sec:ValidCC} presents our evaluation methodology and results.
Section~\ref{sec:discussion} discusses results, lessons learned, and the improvement on the state of practice.
Section~\ref{sec:related} presents related work.
Section~\ref{sec:conclusion} concludes the work.

 \section{Case Study}
\label{sec:caseStudy}
This section presents our automotive case study: it describes the controlled system (\Cref{sec:controlled}) and its functional and drivability requirements (\Cref{sec:Requirements}).

\begin{figure}[t]
    \centering
    \includegraphics[width = 0.8\linewidth]{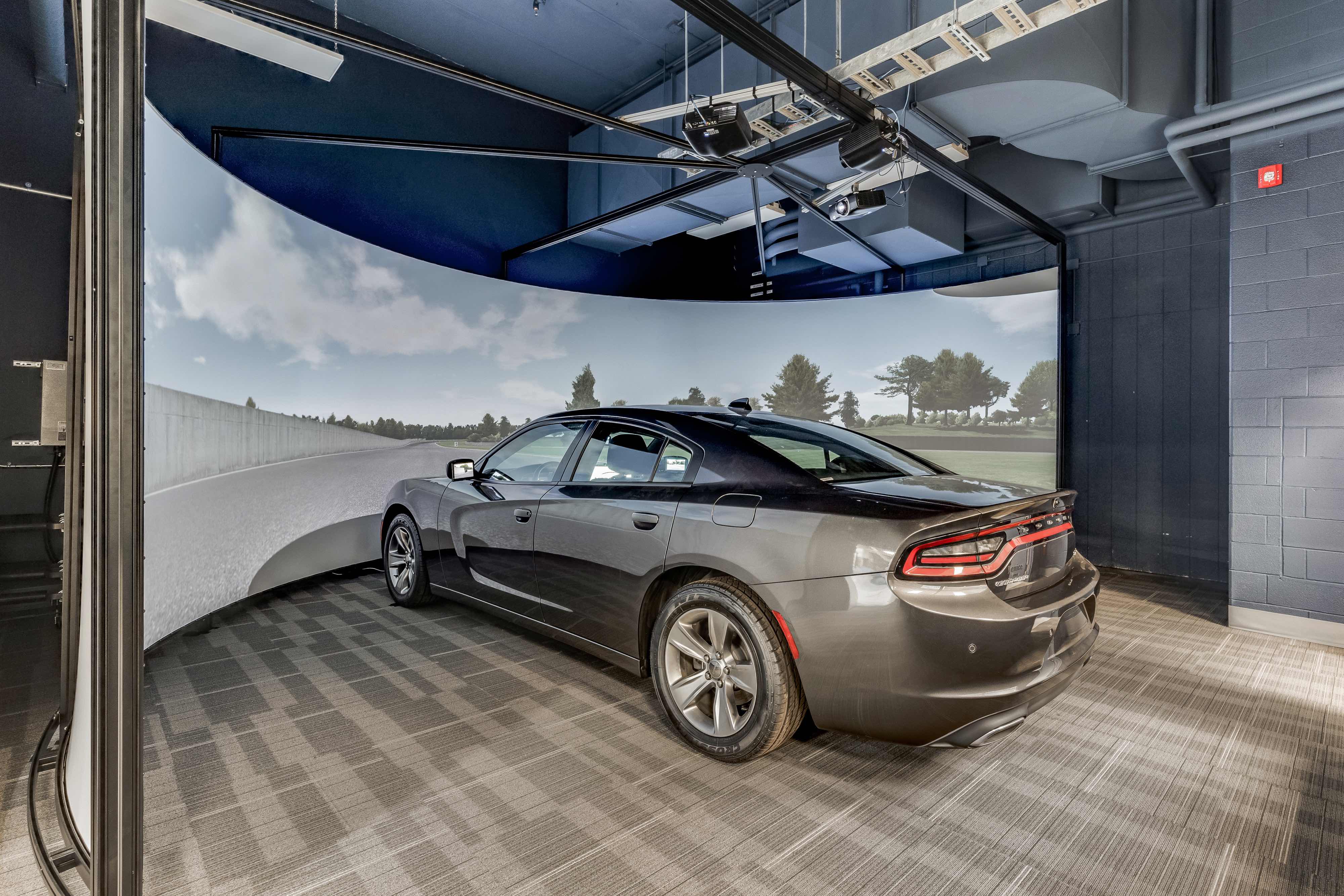}
    \caption{MARCdrive simulator.}\label{fig:MARCdrive}
\end{figure}

\begin{figure}[t]
    \centering
    \includegraphics[width =1.0 \linewidth]{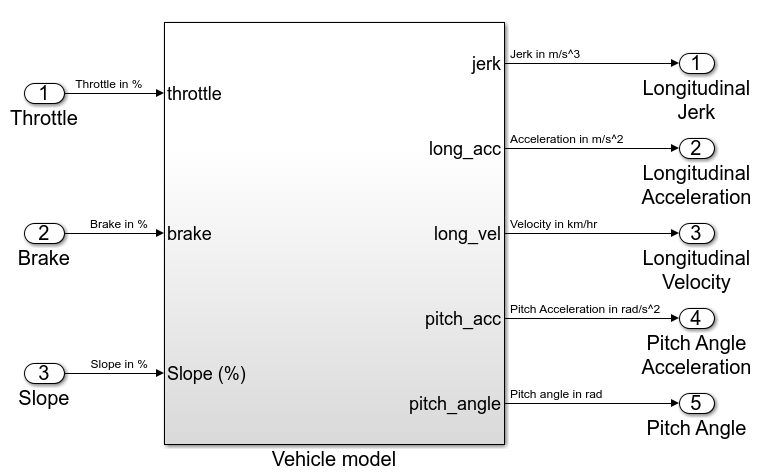}
    \caption{\simulink model under analysis.}
    \label{fig:SimulinkModel}
\end{figure}

\subsection{Controlled System}
\label{sec:controlled}
Our case study uses the VI-CarRealTime (VICRT) Simulator~\cite{vicar}. 
The controlled system is a vehicle traveling on the road: 
the inputs and outputs of the vehicle relevant to our study are schematized in \Cref{fig:SimulinkModel}.
The inputs are the throttle, brake, and slope percentages that  respectively assume values within the ranges $[0,100]$\%, $[0,100]$\%, and $[-5,5]$\%.
The outputs are the longitudinal jerk, acceleration, and velocity, and the pitch angle and acceleration.
The goal of the software engineer is to design a cruise controller (CC) that measures the longitudinal velocity, the pitch angle, and slope percentages and acts on the throttle and brake to ensure the satisfaction of the functional and drivability requirements (see~\Cref{sec:Requirements}).

\begin{figure}[t]
    \centering
\includegraphics[width = 0.7\linewidth]{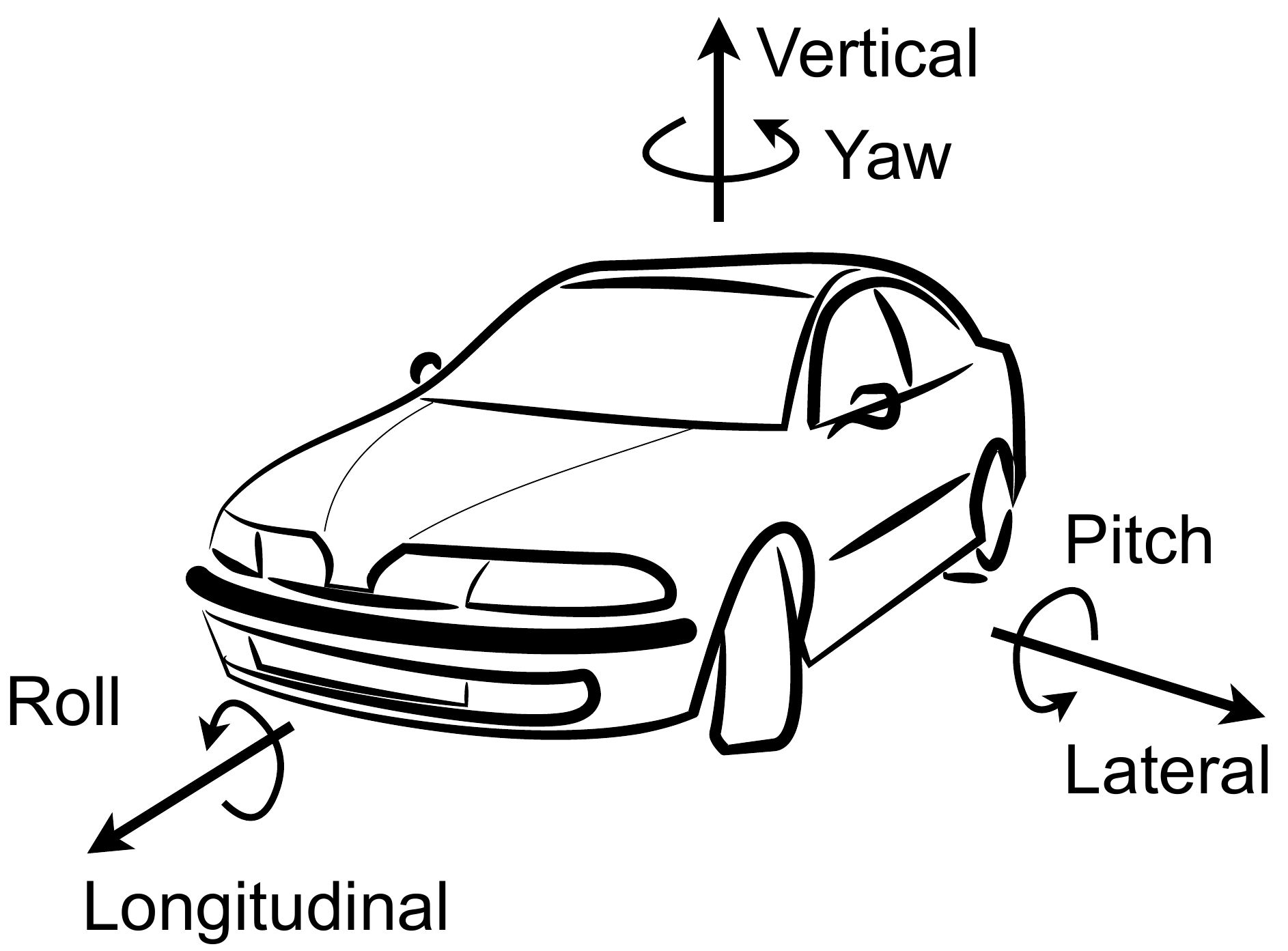}
    \caption{Axes used for the definition of the drivability requirements.}\label{fig:seated}
\end{figure}

During the CC development, the software engineer can use SBST to automatically search for failure-revealing test cases, i.e., test cases that show a violation of the functional and drivability requirements.
As we will describe later (\Cref{sec:ValidCC}),  we performed software-in-the-loop (SIL) and hardware-in-the-loop (HIL) simulations to test the CC model. 
For SIL simulations, the vehicle  is represented by a \simulink model provided by the VICRT framework. 
This model contains \VICRTBlocks \simulink blocks and relies on an XML file produced by VICRT.
The XML file contains all the vehicle properties (e.g. motor speed-torque curves, the tire properties), used by the \simulink model to simulate faithfully the vehicle dynamics.

For HIL simulations, we considered the MARCdrive simulator
~\cite{marcdrive}
from the McMaster Automotive Resource Centre (MARC)~\cite{MARC} shown in \Cref{fig:MARCdrive}.
This simulator integrates VICRT with the model of a Sedan Car, i.e., a passenger car with separate compartments for the engine, passengers, and cargo.
It is a high-fidelity, customizable simulator that enables engineers to test their projects. 
The projections of the road and surrounding environment on the 210-degree screen, the sound system, and the active steering wheel, brakes, driver’s seat, and seatbelt enable the driver to perceive the vehicle dynamics (speed and acceleration). 
The VICRT calculates the dynamics of the car and provides feedback to the driver.

For our experiments, we considered the vehicle traveling on a straight trajectory.
We analyzed this trajectory since it is commonly considered for acceleration tests by 
automotive CC developers and vendors~\cite{Bruck2022,AccellerationTest}.

\subsection{Functional and Drivability Requirements}
\label{sec:Requirements} 
We analyzed the functional (F1) and  drivability requirements (D1, D2, D3) from \Cref{tab:my_label}.
These requirements are largely considered in the research literature; the interested reader can consult the following publications: 
\cite{shaout1997cruise} for F1, \cite{muller2013can,hoberock1976survey,schuette2005hardware} for D1,
\cite{takahama2018model} for D2,
and \cite{Bruck2022} for D3.

The functional requirement F1 requires the vehicle speed to reach and stay within 3 km/hr of the desired velocity after 30 seconds of the velocity being set.
We selected the threshold of 30 seconds as it allows for an operating range of 100 km/hr for the CC while still meeting all drivability requirements.

The drivability requirements limit the longitudinal acceleration and jerk, and  the pitch acceleration (see \Cref{fig:seated}) within given thresholds ranges.

The drivability requirement D1 requires the longitudinal acceleration to remain within the range $[$\LongAccBack, \LongAccFor$]$ (i.e., $[-0.36g,~0.51g]$).
The threshold values for the longitudinal acceleration were defined after testing different acceleration and braking scenarios on two authors using the MARCdrive simulator.
The selected range includes the expected upper limit of 1.47$\frac{m}{s^{2}}$ for acceptable longitudinal acceleration in public transportation~\cite{hoberock1976survey}.

The drivability requirement D2 requires the absolute value of the acceleration of the vehicle on the pitch axis to not exceed \PitchAcc.
This value was determined experimentally on the MARCdrive simulator, in the same way as the longitudinal acceleration thresholds.
We verified that these values are reasonable by consulting relevant studies in the field (e.g.,~\cite{colombet2017tilt}).

The drivability requirement D3 requires the longitudinal jerk (i.e., the acceleration change ratio over time) to be lower than \JerkThres.
This value is considered comfortable for the driver and passengers~\cite{huang2004fundamental}. 

After defining the requirements of the system, the development of the CC began.

\begin{table}[t]
    \centering
     \caption{Requirements for our controller the  VI-CarRealTime Simulator controller.}
    \label{tab:my_label}
    \begin{tabular}{l p{7.5cm}}
        \toprule
        \textbf{ID} & \textbf{Description}  \\
        \midrule
        F1 & After every change, the system shall reach and stay within \FuncTol of the desired speed after \FuncTime.\\
        D1 & The longitudinal acceleration of the vehicle shall not exceed \LongAccBack \ and \LongAccFor.\\
        D2 & The absolute value of the jerk of the vehicle shall not exceed \JerkThres.\\
        D3 & The absolute value of the acceleration of the vehicle on the pitch axis shall not exceed \PitchAcc. \\
         \bottomrule
    \end{tabular}
\end{table}

\section{Development of Cruise Control}
\label{sec:DevCC}
 The CC model was developed by two of the authors: 
 a mechatronics fourth-year bachelor student, and a first-year Ph.D. student in software engineering with large experience in the development of controllers for CPS. 
The development activity required approximately \developmentTime.
This includes the time required to develop the CC versions and the time needed for testing them.
In this section, we summarize the characteristics of the different versions of the CC and present its final version.
The testing activity will be discussed in \Cref{sec:ValidCC}.

\Cref{tab:vers_ctrl} lists the different versions of the model of the CC. 
For each version of the model, the table provides a description of the changes introduced in that version, the number of inputs, and the number of blocks in the model. Finally, the table shows which requirements were considered during the testing activities (discussed in \Cref{sec:sbst} and \Cref{sec:ValidCC}) on each version: the functional requirement F1 for the first two versions of the model, and the requirements F1, D1, D2, D3 for the remaining versions. 

The model is developed incrementally, and new features are added to the CC until all the requirements are satisfied.
For example, 
version 1.0 is a PID (Proportional, Integral, Derivative) software controller that changes the vehicle's throttle depending on the difference between the velocity and the desired velocity.
The higher the difference, the larger the throttle.
Version 2.0 also acts on the vehicle brakes to reduce the vehicle speed when the desired velocity is lower than the actual velocity.
Version 3.0 adds a smoothing algorithm to smooth non-continuous outputs from the PID controller.

\begin{figure*}
    \centering
    \includegraphics[width=0.95\linewidth]{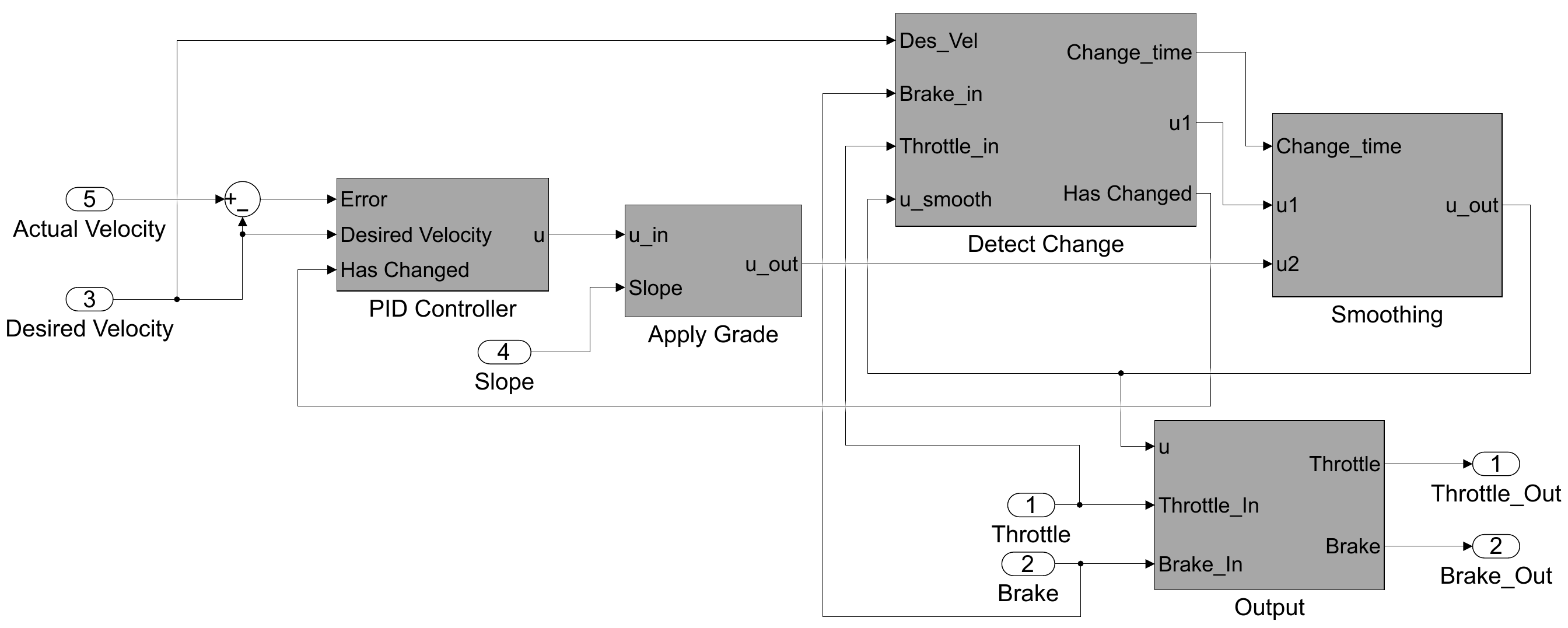}
    \caption{Cruise Controller structure in the latest version.}
    \label{fig:cruiseController}
\end{figure*}

The final version of the CC contains \NBlocksCC blocks. 
\Cref{fig:cruiseController} presents the high-level structure of the final version of the CC. The CC consists of five main subsystems detailed as follows.

\begin{enumerate}
    \item 
\emph{PID Controller.} Calculates the difference between the desired velocity and actual velocity of the system, and outputs the sum of the error, its derivative, and its integral, each multiplied by a certain gain. 
A proportional gain is also added directly to the desired velocity to compensate for different speeds requiring different amounts of throttle to maintain. 
This subsystem also uses changes detected in the \emph{Detect Change} subsystem to reset the integral portion of the controller.
\item 
\emph{Apply Grade.}
Modifies the output of the PID controller depending on the slope of the road: it respectively increases or decreases  depending on whether the slope is uphill or downhill. 
This operation compensates for more or less brake and throttle being necessary for those conditions. 
\item 
\emph{Detect Change.} 
\label{subsec:Detect Change}
Monitors changes in the driving mode that can cause discontinuous behaviors in the CC and takes appropriate actions.
Monitored changes include (a)~the driver releasing the throttle and brake pedals causing the CC to automatically take over, and (b)~the desired velocity changing.
\item
\emph{Smoothing.} 
Creates a smooth transition between the old and the new throttle and brake values after a change in the driving mode.
\item
\emph{Output.} Splits the throttle and brake signals and sends them to the vehicle.
Deactivates the cruise control if either throttle or brake applied by the driver is above 5\% of their maximum value.
\end{enumerate}

SBST was used to check for requirement violations on the different versions of the CC, including its final version.

\begin{table*}[t]
    \centering
    \caption{ID, Description, Requirements ID (Req), Number of Inputs (\#In) and Number of Blocks (\#Bck) of the different versions of our controller.
    }
    \label{tab:vers_ctrl}
    \begin{tabular}{c p{12.5cm} c c c}
    \toprule
         \textbf{ID}  & \textbf{Description} & \textbf{\#In}   & \textbf{\#Bck}$^\ast$ &\textbf{\#Req}\\
         \midrule
         1.0 & Initial version of the PID model for the controller.    &1  &71 &F1\\
         2.0 & Applies brakes when necessary and deactivates the CC by applying the throttle or brake. &1 &94 &F1\\
         2.1 & Introduces drivability requirements. &1 &94 &F1,D1,D2,D3\\
         3.0 & Adds smoothing system by limiting the derivative of outputs from the PID controller. &1 &142 &F1,D1,D2,D3\\
         3.1 & Introduces Test Sequence 2. &1 &142 &F1,D1,D2,D3\\
         3.2 & Introduces Test Sequence 3. &1 &142 &F1,D1,D2,D3\\
         4.0 &  Adapts the velocity to the slope of the road and changes the smoothing system. &2 &261 &F1,D1,D2,D3\\
5.0 & Eliminates the spikes in jerk and improves response in downhill conditions. &2 &348 &F1,D1,D2,D3\\
6.0 &Adds slope detection system within the vehicle, allowing for it to be used within the control system. 
&2 &379  &F1,D1,D2,D3\\
         6.1 & Modifies the proportional gain to consider the slope of the road. &2 &380   &F1,D1,D2,D3\\
         6.2 &Makes smoothing more aggressive and changes the conditions that would trigger it. &2 &402 &F1,D1,D2,D3\\
         6.3 & The signal from the PID controller is smoothed before being applied to throttle or brake. 
&2 &342 &F1,D1,D2,D3\\
         6.4 &Changes the sigmoid smoothing equation to have an initial derivative closer to 0. &2 &335 &F1,D1,D2,D3\\
         6.5 & Uses cubic splines that hit and stay at zero for the throttle and brake profiles. &2 &332 &F1,D1,D2,D3\\
         6.6 &Removes all possibility of brake and throttle signals being on at the same time. &2 &330 &F1,D1,D2,D3\\
         7.0 &Prevents the throttle signal from ever going below 5, which causes a large spike in jerk. &2 &339  &F1,D1,D2,D3\\
         7.1 &Limits the maximum power of brakes preventing negative acceleration violations.  &2 &339 &F1,D1,D2,D3\\
         7.2 &Changes the brake profile to a degree 5 polynomial, making it overall smoother. &2 &329  &F1,D1,D2,D3\\
         7.3 &Resets in the integral in the PID controller to prevent large overshoots after long run times. &2 &337  &F1,D1,D2,D3\\
         7.4 &Introduces Test Sequence 5.  &2 &337 &F1,D1,D2,D3\\
         7.5 &Introduces Test Sequence 6. &2 &337  &F1,D1,D2,D3\\
    \bottomrule
    \multicolumn{5}{l}{* This figure counts the VICRT mask as a single block, though it actually contains \VICRTBlocks blocks.}\\
    \end{tabular}
\end{table*}
 
\section{Search-based Software Testing}
\label{sec:sbst}
This section presents an overview of SBST tools for \simulink models  (\Cref{sec:simulink}), and \NAME~\cite{Hecate}, the framework we selected in this work (\Cref{sec:preparation}).

\tikzstyle{output} = [coordinate]

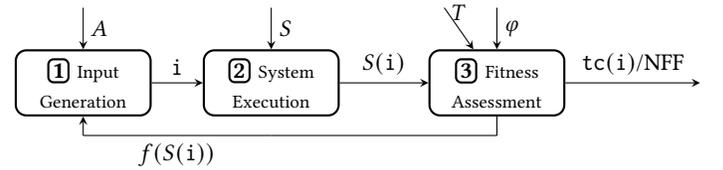
\begin{figure}[t]
\begin{tikzpicture}[auto,
 block/.style ={rectangle, draw=black, thick, fill=white!20, text width=5em,align=center, rounded corners},
 block1/.style ={rectangle, draw=blue, thick, fill=blue!20, text width=5em,align=center, rounded corners, minimum height=2em},
 line/.style ={draw, thick, -latex',shorten >=2pt},
 cloud/.style ={draw=red, thick, ellipse,fill=red!20,
 minimum height=1em}]
 
\draw (0,0) node[block] (Input) {\phase{1} \footnotesize Input \\ Generation};
\node[block, right of=Input,node distance=2.5cm] (Model){\phase{2} \footnotesize System\\  Execution};
 
\node [block, right of=Model,node distance=3cm] (Requirement) {\phase{3} \footnotesize Fitness\\ Assessment};

\node [output, right of=Requirement,node distance=2.7cm] (uoutput) {};
\node [output, above of=Input,node distance=1cm] (Constraints) {};
\node [output, above of=Model,node distance=1cm] (InputModel) {};
\node [output, below of=Requirement,node distance=0.7cm] (v1a) {};
\node [output, left of=v1a,node distance=3cm] (v1) {};
\node [output, below of=Input,node distance=0.7cm] (v2) {};
 \node [output, below of=Requirement,node distance=1.2cm] (mftwo) {};
\node [output, below of=Requirement,node distance=1.2cm] (mf) {};
\node [output, above of=Requirement,node distance=1cm] (ff) {};
\node [output, above of=Requirement,node distance=1cm] (budget1) {};
\node [output, above of=Requirement,node distance=1cm] (budget2) {};
\node [output, left of=budget2,node distance=0.7cm] (budget3) {};

\draw[-stealth] (budget3) -- (Requirement)
    node[midway,above]{$\budget$};

\draw[-stealth] (Input.east) -- (Model.west)
    node[midway,above]{$\inputs$};
\draw[-stealth] (Constraints.south) -- (Input.north)
    node[midway,right]{$\asmpt$};
\draw[-stealth] (InputModel.south) -- (Model.north)
    node[midway,right]{$\system$};

\draw[-] (v1) -- (v2)node[midway,below]{$\fitness(\system(\inputs))$};
\draw[-stealth] (v2) -- (Input.south);
 \draw[-stealth] (Model) -- (Requirement.west) node[midway,above]{$\system(\inputs)$};
 \draw[-stealth] (Requirement.east) -- (uoutput) node[midway,above]{$\texttt{tc}(\inputs)$/NFF};
 \draw[-stealth] (ff) -- (Requirement) node[midway,right]{$\requirement$};
 
   \draw[-] (Requirement) -- (v1a) node[midway,left]{};
 
  \draw[-] (v1a) -- (v1) node[midway,left]{};
 \end{tikzpicture}
\caption{Overview of an SBST framework.}
\label{fig:SBST}
\end{figure}

\begin{figure*}[t]
 \subfloat[Parameterized Test Sequence Block.\label{fig:testSequence}]{
    \centering
    \includegraphics[width=\columnwidth]{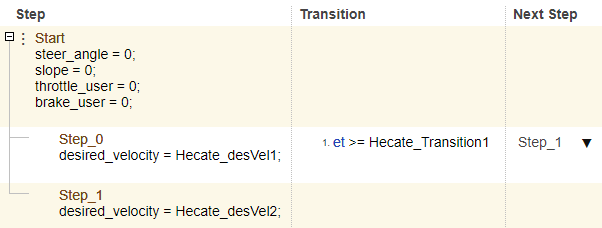}
    \vspace{5mm}
    }
 \subfloat[Test Assessment Block.\label{fig:testAssessment}]{
    \centering
    \includegraphics[width=\columnwidth]{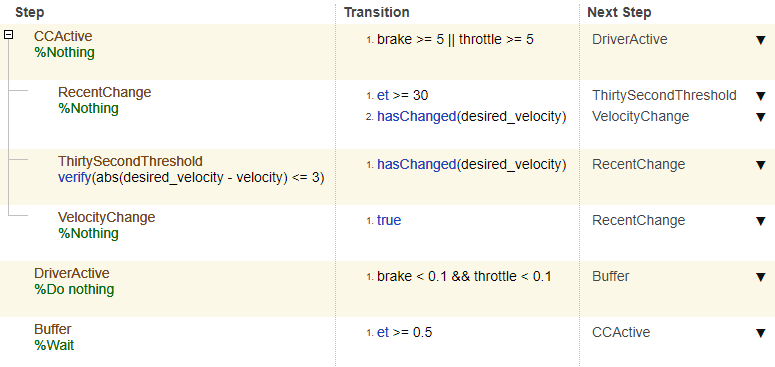}
    }
    \caption{An example of a test sequence and a test assessment block for our cruise controller.}
    \label{fig:tsexample}
\end{figure*}

\subsection{SBST for \simulink Models}
\label{sec:simulink}
The main steps performed by an existing SBST for \simulink models in searching for failure-revealing test cases are depicted in \Cref{fig:SBST}:
squared boxes represent their steps,
arrows connect subsequent steps describing their inputs and outputs, 
arrows with no source and destination are the inputs and outputs of the SBST framework.

\simulink SBST frameworks typically rely on four inputs: 
a model of the system to be tested~(\system); 
an assumption on the system inputs~($\asmpt$), 
a time budget~($\budget$), and a requirement ($\requirement$).
Their output is a failure-revealing test case ($\texttt{tc}(\inputs)$) or an indication that no failure-revealing test case was found (NFF --- No Failure Found) within the time budget.

The SBST frameworks iteratively repeat these steps to search for a failure-revealing test case:
\begin{itemize}
    \item[\phase{1}] \emph{Input Generation}: generates an input~($\inputs$) for the system model~($\system$) that satisfies the assumption~($\asmpt$);
    \item[\phase{2}] \emph{System Execution}: simulates the system model ($\system$) for an input ($\inputs$) and obtains a system execution ($\system(\inputs)$).
    \item[\phase{3}] \emph{Fitness Assessment}: computes the fitness value ($\fitness(\system(\inputs))$) for the system execution ($\system(\inputs)$) and assesses whether the fitness value is below a threshold value. 
\end{itemize}
A test case ($\texttt{tc}(\inputs)$) is failure-revealing if:  (a)~the corresponding input (\inputs) satisfies the assumption ($\asmpt$), and (b)~the fitness value ($\fitness(\system(\inputs))$) is smaller than a threshold value.
Typically, the threshold value is $0$ since the fitness value is negative if the requirement is violated and positive otherwise (e.g.,~\cite{fainekos2006robustness,menghi2019generating,DBLP:conf/arch/ErnstABCDFFG0KM21}).

Several tools for \simulink models implement the framework from \Cref{fig:SBST}, such as 
\ARIsTEO~\cite{Aristeo},
\ATheNA~\cite{Athena},
\Breach~\cite{Breach},
\FalCAuN~\cite{Falstar},
\falsify~\cite{yamagata2020falsification}, 
\FalStar~\cite{ErnstSZH2018FalStar},
\foresee~\cite{falsQBRobCAV2021},
\STGEM~\cite{STGEM},
\STaLiRo~\cite{STaliro},
\PsiTaliro~\cite{PsiTaliro},
and \NAME~\cite{Hecate}.
In this work, we used \NAME~\cite{Hecate} to support the development of our cruise control model since, unlike the other tools, it supports \simulink Test Blocks that are commonly used for testing CPS models.
\vspace{-5mm}
\subsection{\NAME}
\label{sec:preparation}
\NAME supports \simulink Test Blocks (i.e., Test Sequences and Test Assessment blocks).
Specifically, to support SBST, \NAME supports Parameterized Test Sequences~\cite{Hecate}: Test Sequences extended with parameters that are used for automated test case generation. 
Parameterized Test Sequences are detailed later in this section. 
\Cref{fig:tsexample} presents an example of a Parameterized Test Sequence (\Cref{fig:testSequence}) and a Test Assessment (\Cref{fig:testAssessment}) block for requirement F1 of our cruise controller.
The Test Sequence block describes the input of the system; the Test Assessment block describes the requirement to be satisfied.
Both the Test Sequence and the Test Assessment are Test Blocks; therefore, they share part of their syntax.
The syntax and semantics of Test Sequence and Test Assessment blocks are summarized in the following paragraphs. 
An extensive description can be found online~\cite{TestAssessment,TestSequence}.

Test Blocks consist of \emph{test steps} connected by \emph{transitions}.
The Parameterized Test Sequence block in \Cref{fig:testSequence} and the Test Assessment block in \Cref{fig:testAssessment} contains three (\step{Start}, \step{Step\_0},
\step{Step\_1}) and six (\step{CCActive},
\step{RecentChange},
\step{ThirtySecondThreshold},
\step{VelocityChange},
\step{DriverActive},
\step{Buffer}) test steps respectively.
Test steps are hierarchically organized. 
For example, the test step
\step{RecentChange} of \Cref{fig:testAssessment} is nested within the test step
\step{CCActive}. 
At each time instant, the Test Blocks are in exactly one `child' test step and its `ancestors' in the hierarchy.
For example, if the Test Assessment in \Cref{fig:testAssessment} is in the child test step \step{VelocityChange}, it is also in all of its ancestors, namely \step{CCActive}. Whenever a Test Block enters a `non-child' test step, it also enters its first child test step. 
For example, if the
\step{CCActive} test step of \Cref{fig:testAssessment} is entered, its child test step
\step{RecentChange} is also entered.

\emph{Transitions} define how a Test Block switches between test steps. 
They connect a source and a destination test step.
They are labeled with a Boolean formula representing a condition for the transition to be fired.
For example, transitions of the Test Sequence block from \Cref{fig:testSequence} specify how the Test Sequence switches between the test steps \step{Step\_0} and \step{Step\_1}.
Whenever a non-child test step is exited, so are all of its `descendants'. 
For example, if the Test Assessment of \Cref{fig:testAssessment} is in the test step \step{CCActive} and the value of the \simulinkvariable{brake} and the \simulinkvariable{throttle}  become greater than~$5$, the \step{CCActive} test step is \step{RecentChange} is also left.

Test steps from Test Sequence and Test Assessment blocks differ in their purpose. Test Sequences define the values assumed by the input signals. For example, the test step \step{Start} 
of the Test Sequence in \Cref{fig:testSequence} assigns the value 0 to the variable \simulinkvariable{throttle}.
Unlike Test Sequence blocks, Test Assessment blocks contain verification statements, e.g., \lit{verify}, which check whether a logical expression evaluates to \TrueValue  or \FalseValue.
For example, the statement \lit{verify}(\simulinkvariable{abs}(\simulinkvariable{desired\_velocity}-\simulinkvariable{velocity})$<=$3)) 
of the test step \step{ThirtySecondThreshold} 
in \Cref{fig:testAssessment} checks whether the absolute difference between the desired velocity and the velocity of the vehicle is lower than $3$ km/h.

A \emph{test case} is made by a Test Sequence block and a Test Assessment block.
The Test Sequence block generates  a test input, i.e., a set of input signals, that are fed into the \simulink simulator. 
A test input is then executed by simulating the model for the input signals associated with the test sequence. 
The Test Assessment block evaluates if the output signals of the model lead to a violation of the expressions of its \lit{verify} statements. 
A Test Assessment block is typically associated with the requirements of the system.
For example, the Test Assessment block from \Cref{fig:testAssessment} checks whether the conditions from requirement D1 holds.
A test case is failure-revealing if the test input generated from the Test Sequence block leads to a set of output signals that violate the conditions specified by the Test Assessment block. 

Parameterized Test Sequences enable the addition of search parameters to the Test Sequence block. 
These parameters are represented by variable names preceded by the keyword Hecate.
For example, the Test Sequence block from \Cref{fig:testSequence} contains three search parameters: \simulinkvariable{Hecate\_desVel1}, \simulinkvariable{Hecate\_desVel2}, and \simulinkvariable{Hecate\_Transition1}.
These search parameters (detailed in \Cref{tab:parameters}) represent the desired initial velocity, and  final velocity, and the time for which the initial velocity is maintained.

\NAME is an automated SBST framework that implements the framework from \Cref{fig:SBST}.
For the input generation phase (\phase{1}),
\NAME iteratively generates new Test Sequences by assigning values to the parameters of the Parameterized Test Sequences.
For the system execution phase (\phase{2}), \NAME relies on the \simulink simulator to produce the output signals associated with a Test Sequence block.
Finally, for the fitness assessment phase (\phase{3}),  \NAME relies on a specific translation that converts Test Assessment blocks into fitness functions (the interested reader can refer to~\cite{Hecate} for additional information).
Phases (\phase{1}, \phase{2}, and \phase{3}) are executed until a failure revealing test case is detected, or a time budget (number of test cases evaluated) is exceeded. In the first case, \NAME returns the search parameter values associated with the failure-revealing test case.
Otherwise, it returns the NFF value.

We use \NAME  to assess the usefulness of SBST.

\begin{table}[t]
    \caption{Test Sequence Parameters definition and range.}
    \label{tab:parameters}
    \centering
\begin{tabular}{l p{47mm} r r}
    \toprule
    \textbf{Parameter} &\textbf{Description}     &\textbf{Min}   &\textbf{Max}\\
    \midrule
        \emph{des\_vel1}$^{\ast}$    &Desired initial velocity.     &50 &150\\
        \emph{des\_vel2}$^{\ast}$    &Desired intermediary velocity.   &50 &150\\
        \emph{des\_vel3}    &Desired final velocity.      &50 &150\\
        \emph{transition1}  &Time for the initial velocity.  &30 &40\\
        \emph{transition2}  &Time for the intermediary velocity.      &0  &20\\
        \emph{slope}    &The slope of the road.  &-4 &4\\
        \emph{verShift} &Average value of the slope.     &-1 &1\\
        \emph{period}   &Period of the sinusoidal slope.      &30 &200\\
        \emph{horShift} &Horizontal shift of the slope.    &0 &$\pi$\\
    \bottomrule\\
    \multicolumn{4}{m{8.5cm}}{
    \emph{des\_vel1}, \emph{des\_vel2}, \emph{des\_vel3} are in km/h.
    \emph{transition1}, \emph{transition2}, \emph{verShift}, \emph{period}, \emph{horShift} are in s. \emph{slope} and \emph{verShift} are \%.\newline
    $^\ast$  For the first Test Sequence (TS1 in \Cref{tab:TestSequences}), \emph{des\_vel1} and \emph{des\_vel2} are in the range $[0,100]$km/h, since the CC is activated with zero initial velocity.}
    \end{tabular}
\end{table}

\begin{table*}[t]
    \centering
    \caption{Test Sequences considered for testing  the model: Identifier (ID), Number of Steps (\#S), Usage of the Slope Parameter (SP), Manual Startup (MS), Velocity Range (VR), Simulation Time (T), and Description.}
    \label{tab:TestSequences}
    \begin{tabular}{l l l l l l p{13cm}}
        \toprule
        \textbf{ID} &\textbf{\#S}   &\textbf{SP}    &\textbf{MS} &\textbf{VR}   &\textbf{T}  &\textbf{Description} \\
        \midrule
        TS1 &1  &U     &No     &0-100  &70s    &CC starts from stationary and has one step.\\
        TS2 &1  &U     &Yes    &50-150 &70s    &Driver accelerates to a speed above 50 km/h before activating CC with one step.\\
        TS3 &2  &U     &Yes    &50-150 &100s   &Driver accelerates to a speed above 50 km/h before activating CC with two steps.\\
        TS4 &1  &C   &Yes    &50-150 &70s    &Driver accelerates to a speed above 50 km/h before activating CC with one step on a constant slope.\\
        TS5 &2  &C   &Yes    &50-150 &100s   &Driver accelerates to a speed above 50 km/h before activating CC with two steps on a constant slope.\\
        TS6 &2  &S &Yes    &50-150 &100s   &Driver accelerates to a speed above 50 km/h before activating CC with two steps on a sinusoidal slope.\\
        \bottomrule
                \multicolumn{7}{l}{U: Unused, C: Constant, S: Sinusoidal}
    \end{tabular}
\end{table*}

\begin{table}[]
    \centering
    \caption{Parameters used in each Test Sequence.}
    \label{tab:paramvsTS}
    \begin{tabular}{l c c c c c c}
        \toprule
        \textbf{Parameter}  &\textbf{TS1}   &\textbf{TS2}   &\textbf{TS3}   &\textbf{TS4}   &\textbf{TS5}   &\textbf{TS6}\\
        \midrule
        des\_vel1       &\cmark     &\cmark     &\cmark     &\cmark     &\cmark     &\cmark\\
        des\_vel2       &\cmark     &\cmark     &\cmark     &\cmark     &\cmark     &\cmark\\
        des\_vel3       &           &           &\cmark     &       &\cmark     &\cmark\\
        transition1     &\cmark     &\cmark     &\cmark     &\cmark     &\cmark     &\cmark\\
        transition2     &           &           &\cmark     &       &\cmark     &\cmark\\
        slope           &           &           &           &\cmark     &\cmark     &\cmark\\
        verShift        &           &           &           &       &       &\cmark\\
        period          &           &           &           &       &       &\cmark\\
        horShift        &           &           &           &       &       &\cmark\\
        \bottomrule
    \end{tabular}
\end{table}

\section{Evaluation}
\label{sec:ValidCC}
Our evaluation estimates the usefulness of SBST in developing CPS by assessing its support for developing a complex CC model for an
industrial simulator and answering  two research questions:
\begin{itemize}
\item[RQ1:] How \emph{effective} is SBST in producing failure-revealing test cases? (\Cref{sec:effectiveness})
\item[RQ2:] How \emph{efficient} is SBST in producing failure-revealing test cases? (\Cref{sec:efficiency})
\end{itemize}
RQ1 assesses how helpful SBST is in detecting model failures.
RQ2 assesses the time required for detecting these failures.

\subsection{Effectiveness (RQ1)}
\label{sec:effectiveness}
To assess the effectiveness of SBST in producing failure-revealing test cases, we proceeded as follows. 

\textbf{Methodology.} We ran our SBST framework for each version of the CC model from \Cref{tab:vers_ctrl} using the SIL simulations.
We set the cut-off value of \maxiterationsUR iterations: since a single model simulation requires \simulationtime, this value ensures that the test case generation takes approximately 30min (at maximum \maxsimulationtimeUR).
If our test case generation framework returned a failure-revealing test case, we assessed the test case to find and fix the source of the error.
We iteratively modified the CC model, and re-ran the test case generation framework, until the test case generation framework was not able to produce any failure-revealing test case for that requirement.
When the SBST framework was not able to produce any failure-revealing test case, we proceeded by extending the test scenario (e.g., introducing slope as a parameter). In our experiments, the input generation algorithm was Uniform Random \cite{arcuri2011random}, since \maxiterationsUR iterations do not enable other optimization algorithms to converge to a good solution.

We ran our SBST framework by considering six different Test Sequences detailed in \Cref{tab:TestSequences}: as new features were added to the model, we considered more complex Test Sequence blocks to trigger these features. 
For example, the Test Sequence \emph{TS1} (detailed in  \Cref{fig:testSequence}) activates the CC immediately, with an initial vehicle speed of 0 km/h (no Manual Startup).
We considered a more realistic Test Sequence (\emph{TS2}),  which had the driver accelerate to 50 km/hr before activating the CC (with Manual Startup), when the CC could switch itself off when the driver applied brake or throttle.
\Cref{tab:testing_results} reports the Test Sequences considered for the different versions of the model.
The Test Sequences rely on the Test Sequence parameters from \Cref{tab:parameters}.
A Test Assessment block models the requirements of the CC.  
The SBST algorithm searches values for the parameters of the Test Sequences that violate the conditions specified by the Test Assessment blocks, i.e., lead to requirement violations.
If a violation is detected, the corresponding Test Blocks (Test Sequence and Test Assessment) of the failure-revealing test case are returned.
We verified the failure-revealing test cases also on the hardware platform (see \Cref{fig:MARCdrive}) by running them with the HIL simulator.  

We analyzed how many versions of the CC model the SBST framework was helpful by returning a failure-revealing test case. 
We also verified that the HIL and SIL simulations yielded consistent results for the failure-revealing test cases.

\textbf{Results.} \Cref{tab:testing_results} presents our results. The table reports the version of the model (ID) and the Test Sequences (TS\#) considered for that version of the model.
For each version, the table reports the number of iterations (\#IT) required to identify the failure-revealing test case, which requirement was not satisfied, and a description of the identified failure.
The SBST framework returned a failure-revealing test case for \percentageFailure of the versions of our model (\totalFailure out of \totalModelVersions).
The versions of the model that did not yield failing test cases were then tested with \maxiterationsSA iterations of the Simulated Annealing algorithm~\cite{SA}.
Of the models that returned a failing test case, \percentageFailureFunc (\funcFailure out of \totalFailure) were due to functional violations, \percentageFailureDriv (\drivFailure out of \totalFailure) were due to drivability violations, and \percentageFailureFuncDriv (\funcdrivFailure out of \totalFailure) were due to both.

\begin{figure}
    \centering
    \includegraphics[width=0.75\linewidth]{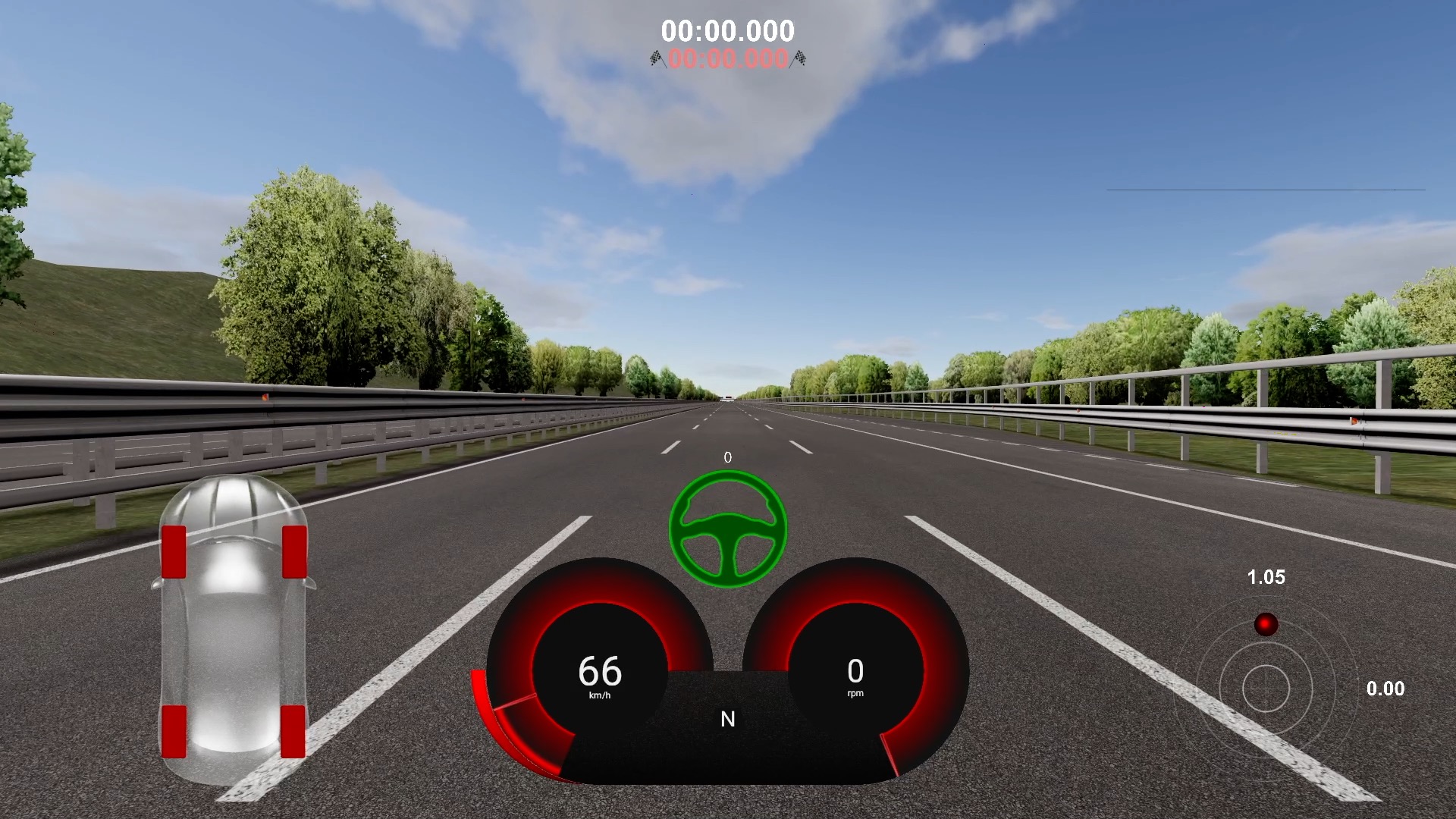}
    \caption{HIL simulation of a test case where the car breaks the drivability requirements. }
    \label{fig:MARCdriveVideo}
\end{figure}

For example, for version 1.0 of the model, the SBST returned a failure-revealing test case for the functional requirement F1. 
The test case revealed that the controller did not act on the vehicle brakes to reduce the speed: when the desired speed was lower than the speed of the car, the vehicle's aerodynamic resistance was insufficient to ensure the vehicle reached the desired speed within $30$s.
The failure was solved by version 2.0 of the model.
For version 2.1 of the model, the SBST returned a failure-revealing test case for the drivability requirement D1. 
\Cref{fig:MARCdriveVideo} shows a snapshot of our simulation, revealing that the CC acts on the brake and the throttle by leading to a longitudinal acceleration of $-11.1\frac{m}{s^2}$ that exceeds the threshold values (\LongAccBack \ and \LongAccFor) specified by the requirement D1.
The corresponding video is available online~\cite{Material}. 

\begin{table*}[t]
    \caption{Model failures identified by analyzing the failure-revealing test cases.}
    \label{tab:testing_results}
    \centering
    \begin{tabular}{c c c  c c c c p{11.5cm}}
    \toprule
      \textbf{ID} & \textbf{TS\#} & \textbf{\#IT} &\textbf{F1} & \textbf{D1} & \textbf{D2} & \textbf{D3} & \textbf{Description of Failure}\\
         \midrule
         1.0 &1 &1
         & \xmark &N.T. &N.T. &N.T.
         &The controller does not act on the brakes to decrease the  speed of the vehicle.\\
         2.0 &1 &$>20$
         & \cmark &N.T. &N.T. &N.T.
         &No failure; introduced drivability requirements.\\
         
         2.1 &1 &9 
         & \cmark & \xmark & \xmark & \cmark
         &The controller  discontinuously applies brakes or throttles. \\

         3.0 &1 &$>20$
         & \cmark & \cmark & \cmark & \cmark
         &No failure detected; introduced test sequence 2.\\
         
         3.1 &2 &$>20$
         & \cmark & \cmark & \cmark & \cmark
         &No failure detected; introduced test sequence 3.\\
         
         3.2 &3 &$>20$
         & \cmark & \cmark & \cmark & \cmark
         &No failure detected; introduced slope parameter.\\
         
         4.0 &4 &1
         &\xmark &\cmark & \xmark &\cmark
         &Numerical errors in the VICRT mask cause random spikes in jerk during run time.\\
         
         5.0 &4 &1
         &\xmark &\cmark &\cmark &\cmark
         &Large negative/positive slopes prevent the vehicle from timely reaching the desired velocity.\\

         6.0 &4 &1
         &\xmark &\xmark &\cmark &\cmark
         &Large changes in desired velocity cause the vehicle to violate acceleration requirements.\\

         6.1 &4 &8
         &\xmark &\cmark &\xmark &\cmark
         &Fails to solve the previous problems.\\

         6.2 &4 &2
         &\cmark &\cmark &\xmark &\cmark
         &Fails to solve the previous problems.\\

         6.3 &4 &1
         &\xmark &\cmark &\xmark &\cmark
         &Fails to solve the previous problems.\\
 
         6.4 &4 &1
         &\xmark &\xmark &\xmark &\cmark
         &Fails to solve the previous problems.\\

         6.5 &4 &2
         &\cmark &\xmark &\cmark &\cmark
         &Fails to solve the previous problems.\\

         6.6 &4 &2
         &\xmark &\cmark &\cmark &\cmark
         &Fails to solve the previous problems.\\

         7.0 &4 &7
         &\xmark &\cmark &\cmark &\cmark
         &Large negative changes in desired velocity are impossible to reach without violating D1.\\

         7.1 &4 &14
         &\xmark &\cmark &\cmark &\cmark
         &Very specific test cases marginally fail functional requirements.\\

         7.2 &4 &4
         &\xmark &\cmark &\cmark &\cmark
         &Fails to solve the previous problem.\\
         
         7.3 &4 &$>20$
         &\cmark &\cmark &\cmark &\cmark
         &No failure detected; introduced test sequence 5.\\

         7.4 &5 &$>20$
         &\cmark &\cmark &\cmark &\cmark
         &No failure detected; introduced test sequence 6.\\

         7.5 &6 &$>20$
         &\cmark &\cmark &\cmark &\cmark
         &No failure detected.\\
         
    \toprule

    \end{tabular}\\
    \xmark: Requirement not met;
    \cmark: Requirement met;
    N.T.: Requirement not tested.    
    
\end{table*}

\begin{Answer}[RQ1 - Effectiveness]
The answer to RQ1 is that, for our automotive model, SBST returned a failure-revealing test case for \percentageFailure of our model versions (\totalFailure out of \totalModelVersions).
The failure-revealing test cases enabled the identification of the failures described in \Cref{tab:testing_results}.
The test cases revealed a violation of the drivability requirements for \percentageFailureDriv of the cases.
\end{Answer}

\subsection{Efficiency (RQ2)}
\label{sec:efficiency}
To assess the efficiency of our SBST framework in producing failure-revealing test cases, we proceeded as follows.

\begin{figure}[t]
    \centering
    \begin{subfigure}[b]{0.45\columnwidth}
        \centering
        \includegraphics[width = \linewidth]{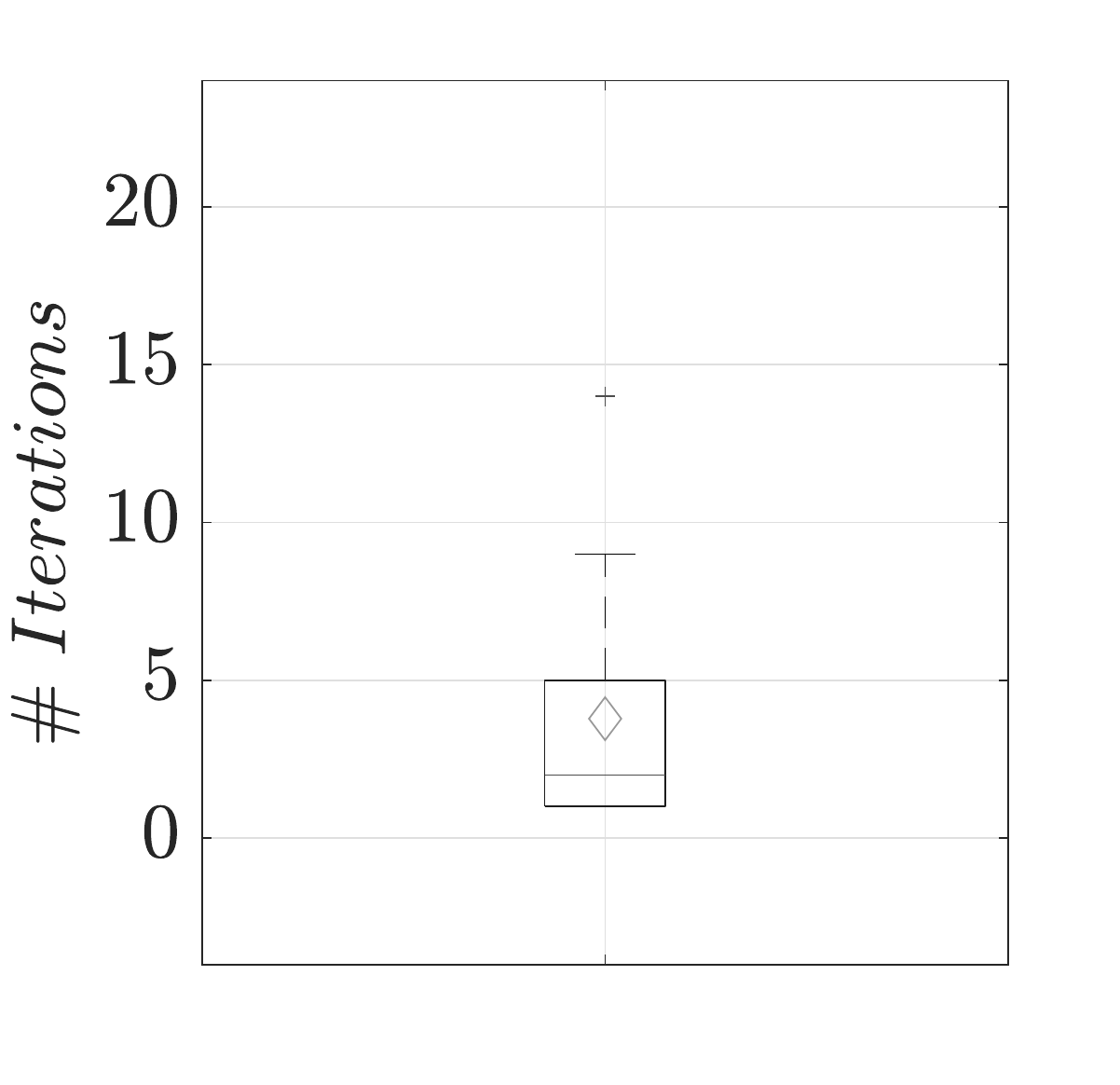}
        \caption{Number of iterations.}
        \label{fig:boxplotIter}
    \end{subfigure}
    \hfill
    \begin{subfigure}[b]{0.45\columnwidth}
        \centering
        \includegraphics[width = \linewidth]{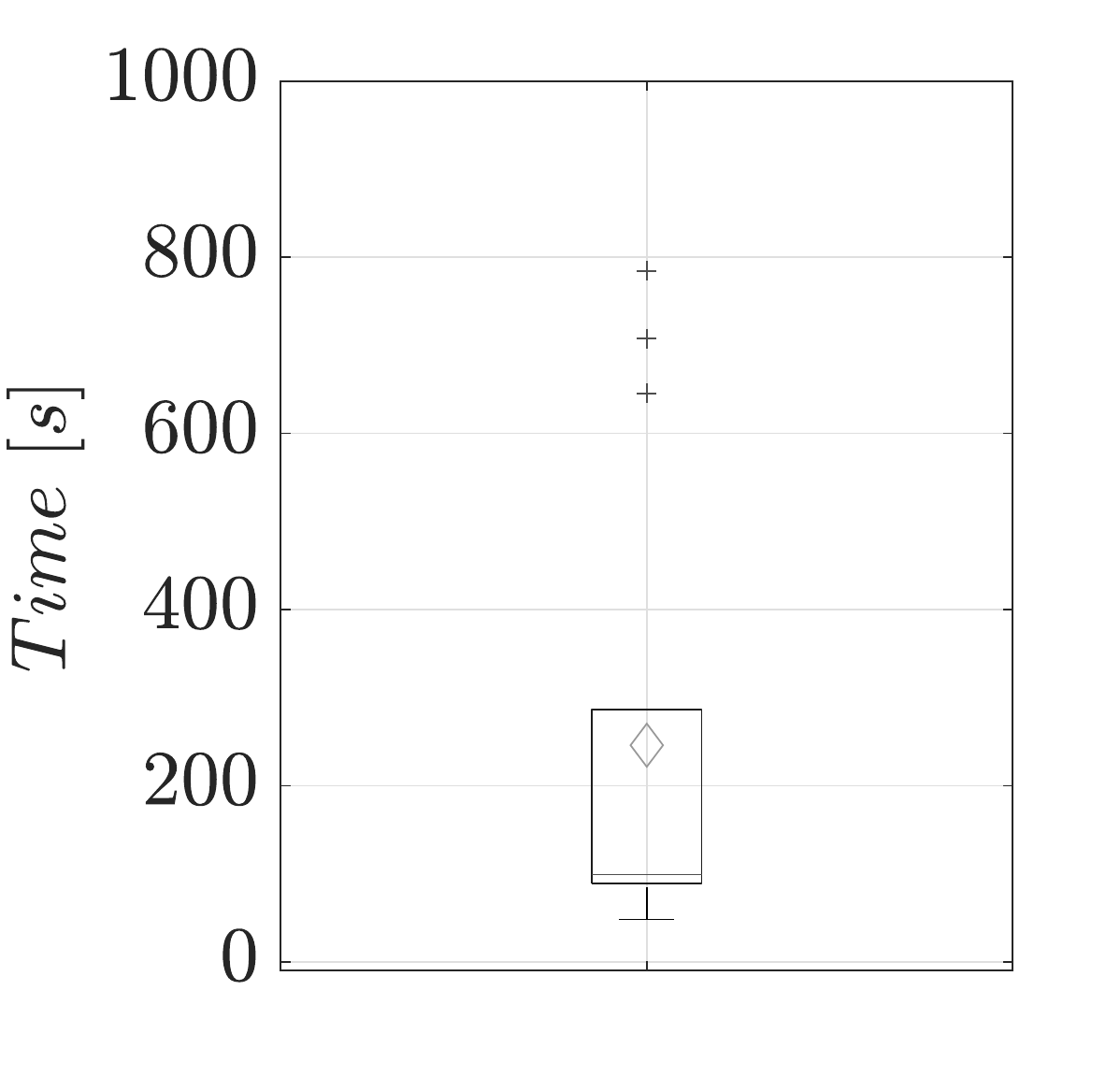}
        \caption{Time.}
        \label{fig:boxplotTime}
    \end{subfigure}
    \caption{Iterations and time of our SBST framework.}
    \label{fig:boxplots}
\end{figure}

\textbf{Methodology.} 
We considered the runs from RQ1 returning a failure-revealing test case.
We analyzed the number of iterations and time required to detect the failure-revealing test case.

\textbf{Results.} 
The box plots in \Cref{fig:boxplots} report the number of iterations (\Cref{fig:boxplotIter}) and the time (\Cref{fig:boxplotTime}) required to produce the failure revealing test cases.
Our SBST framework required, on average \itavg
(\textit{min}=\itmin,
\textit{max}=\itmax,
\textit{StdDev}=\itstd) iterations
 and  \timeavg  (\textit{min}=\timemin,
\textit{max}=\timemax,
\textit{StdDev}=\timestd) to generate these test cases. 

\phantom{x}\\

\begin{Answer}[RQ2 - Efficiency]
The answer to RQ2 is that, for our automotive model, the SBST framework 
required  on average 
\itavg iterations
(\textit{min}=\itmin,
\textit{max}=\itmax,
\textit{StdDev}=\itstd) and
\timeavg  (\textit{min}=\timemin,
\textit{max}=\timemax,
\textit{StdDev}=\timestd) to return the failure-revealing test cases.
\end{Answer}

\subsection{Discussion and Threats to Validity}
Our results confirm the usefulness of SBST in detecting failure-revealing test cases of a complex CC for an industrial simulator (\textbf{P1}): the SBST returned a failure-revealing test case for the majority (\percentageFailure) of our model versions in practical time (\timeavg$\approx4$min).
They also confirm the usefulness of SBST in detecting failure-revealing test cases for drivability requirements (\textbf{P2}):  \percentageFailureDriv of the failure-revealing test cases showed a violation of a drivability requirement.
Finally, our results confirm the usefulness of SBST driven by \simulink Test Blocks (\textbf{P3}): we could express all the requirements of the CC using the Test Assessment blocks, and our search space using Parameterized Test Sequences.

Our results are subject to the following threats to validity.

The automotive benchmark model, requirements, and controller versions we analyzed could threaten our results' \emph{external validity} since different benchmark models, requirements, and controllers can lead to different results for the effectiveness and efficiency of the SBST.
The fact that our industrial model is used by many automotive companies (e.g., Brembo~\cite{Brembo}) and taken from the VI-CarRealTime simulator, the functional and drivability requirements come from the research literature (e.g.,~\cite{shaout1997cruise}, \cite{muller2013can,hoberock1976survey,schuette2005hardware}, \cite{takahama2018model}, \cite{Bruck2022}), and the CC model  is a large and complex model developed in eight months mitigate this threat. The number (\totalBlocks) of blocks of the final model (system under test plus CC) is a reasonable approximation to a small to mid-sized realistic industrial model~\cite{boll2021characteristics}.

The values of the configuration parameters of our SBST could threaten the \emph{internal validity} of our results: a higher number of iterations for our SBST framework or a different search algorithm can produce different results~\cite{4983313}.
To mitigate this threat, for the versions of the CC for which the SBST did not return any failure-revealing test case, we increased the number of iterations to \maxiterationsSA and considered an additional search algorithm (Simulated Annealing~\cite{SA}). 
This configuration of our SBST framework also did not return any failure-revealing test case for these versions of the CC.

\section{Discussion}
\label{sec:discussion}
This section reflects on the lessons learned (\Cref{sec:lessons}), the improvement on the state of the practice (\Cref{sec:improvement}), and the generality of our results (\Cref{sec:generality}). 

\subsection{Lessons Learned}
\label{sec:lessons}
The results of our experimental evaluation confirm
the usefulness of SBST in detecting failure-revealing test cases of a complex CC~(\textbf{P1}), for drivability requirements~(\textbf{P2}), and \simulink Test Blocks~(\textbf{P3}).

In addition, we learned many lessons from our experimental evaluation: we report on five of these lessons (L1-L5).

\emph{Running many simulations  is compute-intensive}~(L1). Testing our automotive case study requires significant computational time.
For our automotive system, executing a test case takes around \simulationtime. Therefore, running \maxiterationsUR iterations of our SBST framework takes approximately \maxsimulationtimeUR.  
This time may be prohibitive for many applications that require executing a vast number of test cases that need to be executed in-house for confidentiality reasons or due to specific configurations of the machines that are difficult to replicate on large computing platforms. 

\emph{The input domain of our model is large} (L2). 
 In our testing activity, we considered the vehicle traveling in a straight trajectory. We also assumed that no other vehicles were present on the road, and the CC can only be set to speeds between 50 km/h and 150 km/h.
 Despite these assumptions, the input domain is infinite: 
 the values of the slope percentage (represented as a real variable) can be changed (across the simulation time) in an infinite number of ways.
In addition, even though our assumptions are reasonable for our experiments, large testing campaigns may require considering other input configurations (e.g., different input shapes), leading to additional problems for the applicability of SBST. 
 The large input domain poses the engineers with trade-off decisions: on the one hand, the engineer wants to increase the number of simulations to increase the number of inputs that are analyzed; on the other hand, since the automotive model is compute-intensive (L1), a limited number of simulations can be executed.

\emph{HIL testing is laborious} (L3). 
HIL requires time and effort.
Configuring our hardware platform for our experiments required significant effort.
Starting and shutting down the simulator requires 10 minutes and the channels to transmit the signals to and from the hardware must be checked manually for each model version (24 input signals and 21 output ones).
Running experiments using the hardware platform may also lead to unpredicted problems: for example, while running our experiments, the light of one of the projectors of our simulator stopped working, and we had to replace it.
Replacing the light of the projector required a few weeks.
In addition, HIL testing activities could only be performed within the McMaster facilities to access the hardware platform, and we had to ensure the presence of humans monitoring the simulator's behavior as mandated by our laboratory regulations.
The presence of a human limits the number of HIL tests that can be executed on the hardware platform that, therefore, need to be carefully selected.

\emph{Results of HIL and SIL simulations may differ} (L4). 
We notice discrepancies in the results of HIL and SIL simulations for some of the test cases. 
Specifically, HIL simulations introduce some noise (with non-zero average) on the throttle pedal, brake pedal, and steering wheel angle signals that were not present in the SIL simulations.
Furthermore, the HIL simulations have a sampling time of $10ms$, while SIL uses $1ms$.
This causes a difference in the way the jerk is computed and produces much lower values for the HIL results.
These minor discrepancies are enough to change the test verdict in some of the test cases.
We will carefully investigate the discrepancies between HIL and SIL simulations and how to mitigate them in future work.

\emph{SIL testing activities need to be automated} (L5). 
Manually executing the SBST framework after any change in the model is laborious and needs to be automated.
Integrating the SBST activity in a continuous pipeline that automatically fetches the model and runs the SBST activity after each version of the model is produced can alleviate this challenge. 

The challenges we identified in our automotive case study are general and apply to other CPS domains, as argued in the following.

\subsection{Generality of the Results}
\label{sec:generality}
To assess the generality of our results and lessons learned, we critically compared them to those from the literature.

For the usefulness in detecting failure-revealing test cases for a complex CC~(\textbf{P1}),
we do not expect that the number of model versions in which the SBST returned a failure-revealing test case will be the same for other CC, automotive models, and CPS. 
However, we believe that these findings apply to other types of CPS.
Empirical studies from the literature confirm this hypothesis in other domains (e.g., space~\cite{Aristeo}, automotive~\cite{fainekos2012, stocco2023model}, biomedical~\cite{pacemaker}, 
medical~\cite{pacemaker,10.1007/978-3-031-04673-5_5,arcaini2016model}). 

For the usefulness in detecting  failure-revealing test cases for drivability requirements~(\textbf{P2}),
we do not expect that the percentage of test cases that reveals a violation of the drivability requirements will match the one obtained by analyzing other CC models since the results of SBST strongly depend on the model that is analyzed.
However, we believe that  SBST will help detect failure-revealing test cases for the drivability requirements of other CC models.
Future empirical studies will confirm or refute our hypothesis. 

For the usefulness in detecting failure-revealing test cases for \simulink Test Blocks~(\textbf{P3}), we do not expect the values for the effectiveness and efficiency to match the one obtained by analyzing other systems, since the results depend on the model and the Test Blocks considered by the SBST framework. 
However, we believe that the SBST framework is also useful to analyze \simulink Test Blocks that refer to other CPS.
The results provided by \NAME~\cite{Hecate} confirm our hypothesis.
The authors of this work have shown that SBST was useful in detecting failure-revealing test cases for a large set of benchmarks consisting of 16 \simulink models from different domains (automotive, energy, biomedical, avionics, domotics), including three developed by Lockheed Martin (EU, NNP, TUI)~\cite{mavridou2020ten}.

We believe that the lessons we learned (L1, L2, L3, L4, and L5) generalize to other types of CPS.
There are several additional studies from the literature confirming these hypotheses (i.e.,\cite{Aristeo,Elevator} for L1,
~\cite{koschi2019computationally,laurent2022parameter} for L2, and
~\cite{mehra2015adaptive,9869302} for L3 and L4).

We conclude that our results are likely generalizable and may also apply to other CPS domains. Additional empirical studies and experimental analysis may confirm or confute our hypothesis across the different domains. In the following section, we will also argue that our results significantly improve the state of the practice.

\subsection{Improvement on the State of the Practice}
\label{sec:improvement}
Our results significantly improve the state of practice. 

\emph{Our results show that SBST effectively supports the development of a complex CC for an industrial simulator}~(\textbf{P1}). Our results improve the state of practice by providing empirical data, evidence, and reflections related to using SBST to develop CC.
Our findings, i.e., the percentage of model versions in which the SBST could return a failure-revealing test case and the number of iterations and time needed to compute them, provide automotive companies with empirical data that can help them decide when and how to use SBST and estimate the resources (e.g., time and computational resources) needed to conduct this activity.
    
\emph{Our results show that SBST effectively supports the analysis of drivability requirements}~(\textbf{P2}).
We argued (\Cref{sec:Intro}) that drivability requirements are critical for automotive applications. 
Despite their importance, the research literature does not extensively assess the usefulness of SBST frameworks for these requirements.
This work evaluates the usefulness of SBST for drivability requirements experimentally via a case study supported by an industrial simulator.
Our results improve the state of practice by showing that SBST effectively finds failure-revealing test cases for drivability requirements. 
They also show that the test cases revealed a violation of the drivability requirements for \percentageFailureDriv of the failure-revealing test cases.
This result suggests that automotive industry needs to extensively consider these requirements since, for a significant number of cases, the failure-revealing test cases refer to these requirements rather than the functional requirements of the CC.
This finding will impact the automotive domain, supporting a more extensive adoption of SBST frameworks driven by drivability requirements.

\emph{Our results show that SBST driven by Test Blocks is useful}~(\textbf{P3}). 
Although existing work has shown the effectiveness of SBST for \simulink models, the only framework that supports Test Blocks is 
 \NAME.
It was evaluated by considering a benchmark consisting of 16 \simulink models from different domains.
However, this work does not assess SBST for the end-to-end development of a CPS, i.e., the benchmark does not include a model for which all the versions produced during the software development are available.
Unlike the work introducing \NAME, we assessed the effectiveness of SBST driven by Test Blocks during the end-to-end development of a CC for an automotive system supported by an industrial simulator.
Our results improve the state of practice by showing that SBST driven by Test Blocks effectively supports developer activities and paves the way for adopting this framework within the industry and for more empirical studies assessing the advantages and limitations of the usage of SBST driven by Test Blocks in the end-to-end development of CPS.

  \section{Related Work}
\label{sec:related}
Several works assessed the usefulness of SBST in detecting failure-revealing test cases for the development of CPS (e.g.,~\cite{10.1007/978-3-031-04673-5_5,arcaini2016model,ahmed2020automated,Elevator,10114597,ling2023benefits}) and CCs (e.g.,~\cite{han2022elevator,xu2022uncertainty}). 
Unlike these works, we assessed the usefulness of SBST by considering the end-to-end development of the CC and used an industrial simulator for our simulation.
 
Several other studies have used the VI-CarRealTime industrial simulator.
For example, this simulator was used to analyze alternative steering mechanisms~\cite{10003308}, the relationship of steering and stability~\cite{9980968}, and other aspects of automated driving~\cite{8883327, 8893027, 8920504, 10.1007/978-3-030-31154-4_11, 10003336}. However, none of these studies used the simulator to assess the usefulness of SBST.

Although drivability requirements are extensively considered during the design of CC (e.g.,~\cite{althoff2020provably,opila2011energy,Bruck2022,barroso2023driver,9490149,drivability}), the assessment of SBST techniques typically focuses on the functional requirements of the CC (e.g.,~\cite{8952365,koschi2019computationally,7795751}).
Our results showed the importance of considering the drivability requirements, i.e., for a significant number  of the generated failure-revealing test cases (\percentageFailureDriv), the failure caused a violation of the drivability requirements. 

Finally, based on our knowledge, the only work that assesses the usefulness of SBST driven by \simulink Test Blocks is the one proposed by \NAME~\cite{Hecate}.

 \section{Conclusion}
\label{sec:conclusion}
We assessed the usefulness of SBST in generating failure-revealing test cases for the functional and drivability requirements of a CC of a vehicle.
We used an SBST framework during the end-to-end development of a complex CC developed using \simulink.
We performed SBST relying on a framework that supports \simulink Test Blocks.
We performed both SIL and HIL testing using an industrial simulation environment widely used in the automotive domain.

The empirical results of our experimental evaluation confirm that SBST is useful in detecting failure-revealing test cases of a complex CC, for drivability requirements, and \simulink Test Blocks.
We presented the lessons we learned from our experimentation, reflected on the generality of our results, and discussed how our results improve the state of practice. 
 
\section*{Data Availability}
A replication package containing all of our data, test results, and videos recorded during HIL simulations is publicly available~\cite{Material}.

\bibliographystyle{ACM-Reference-Format}

\end{document}